\documentclass[pre,final,showpacs,showkeys]{revtex4}
\usepackage{graphics}
\usepackage{graphicx}

\newcommand{\be}{\begin{equation}}
\newcommand{\ee}[1]{\label{#1} \end{equation}}
\newcommand{\ba}{\begin{eqnarray}}
\newcommand{\ea}[1]{\label{#1} \end{eqnarray}}
\newcommand{\nl}{\nonumber \\}

\newcommand{\pd}[2]{ \frac{\partial {#1}}{\partial {#2}}  }
\newcommand{\re}[1]{(\ref{#1})}

\newcommand{\exv}[1]{ \langle {#1} \rangle }

\usepackage{amsmath}

\begin{document}

\title{{\bf Zeroth Law compatibility of non-additive thermodynamics\\ }}

\author{ 
  T. S. Bir\'o\(^1\)\  and   P. V\'an\(^{1,2}\)
}
\affiliation{
  { \(^{1}\)Dept. of Theoretical Physics, KFKI Research Institute for Particle and Nuclear Physics,} \\
  { H-1525 Budapest, P.O.Box 49, Hungary} \\
 {\(^{2}\)Dept. of Energy Engineering, Budapest Univ. of Technology and Economics},
  Bertalan Lajos u. 4-6, H-1111, Budapest, Hungary}



\date{\today}

\begin{abstract}
        Non-extensive thermodynamics was criticized among others by stating that the
        Zeroth Law cannot be satisfied with non-additive composition rules.
        In this paper we determine the general functional form of those non-additive composition rules 
        which are compatible with the Zeroth Law of thermodynamics. We find that this
        general form is additive for the formal logarithms of the original quantities
        and the familiar relations of thermodynamics apply to these.
        Our result offers a possible solution to the longstanding 
        problem about equilibrium between extensive and non-extensive systems or systems with
        different non-extensivity parameters.
\end{abstract}

\maketitle
\normalsize


\section{Introduction}


Since Gibbs's famous formula, 
\be
S = \sum_i p_i \ln  \frac{1}{p_i},
\ee{Boltzmann_Formula}
several expressions have been proposed
and are in use for the entropy. Many applications stem from
the field of theoretical informatics \cite{Arn01b,Taneja}. 
A common motivation in the hunt for further formulas is to consider entanglement
or correlation between elementary events or states of the system under 
study. A fairly well known suggestion is due to 
Alfr\'ed R\'enyi; in his formula a parameter, nowadays denoted by $q$,
occurs \cite{Ren61a}.

A more recent endeavour was initiated by Constantino Tsallis \cite{Tsa88a,Tsa09b},
namely to build thermodynamics and statistical physics on a mathematical
formulation counting for the interdependence between
alternative states of a complex system, but possibly remaining closely as simple
and powerful as the classical theory. Based on the non-additive nature of the entropy formula
promoted by Tsallis \cite{Tsa88a,HavCha67a} almost all familiar properties of
the classical thermodynamics have been questioned, re-iterated and are
partially still open. One of the open questions relates the Maximum Entropy Principle
to the Zeroth Law of thermodynamics: How can the
transitivity property of thermal equilibrium and the definition of the
empirical temperature be preserved in this more complex approach? 
It was, in particular, a sharply formulated critique
against using another than the classical Boltzmannian formula, that 
temperature, heat exchange, probability factorizability were not clearly conceptualized and   
in particular that the parameter $q$ could not be interpreted in a way compatible to the Zeroth Law \cite{Nau03a}.

Quite a few attempts were devoted to the clarification of this issue  
\cite{Abe99a,AbeEta01a,MarEta01a,OuChe06a, Tor03a,Sca10a,Abe2001}. 
The generalization of thermostatistics considering non-additive entropy functions has been 
proposed long ago \cite{Tsa88a}, but its relation to alternative suggestions
\cite{Verda,AczelDaroczy,Tanejacikk,Kaniadakis} is not yet completely understood \cite{Tsa09b}. 
One challenge to thermodynamics using non-additive entropy functions is its surmised 
non-compatibility with the Zeroth Law. 
On the other hand, it was  shown recently, that one may achieve some consequences 
of these theories - most prominently power-law tailed equilibrium distributions of the energy 
in canonical statistical systems - assuming that the entropy is additive, but other  
fundamental thermodynamic variables, e.g. the energy, are not \cite{BirPes06a,Bir08a}.  
Moreover, there are indications, that both entropy and energy may be non-additive \cite{Wan02a,Sca10a}.

In this article we approach the problem of non-extensive equilibrium in a more general setting: 
given - as a rule non-additive -
composition rules for the classical extensives, originally treated as additive
quantities, what constraints can be derived from the Zeroth Law for their
functional form? What consequences can be, furthermore, formulated for
the usefulness of one or another entropy formula, and in particular what
is the role of any $q$-like parameter in the thermodynamical equilibrium state?


\section{The Zeroth Law: factorizability}

In order to arrive at answers to the questions raised in the Introduction, 
we briefly repeat the main steps of the classical construction of the Zeroth Law and then
we generalize it to non-additive composition rules of entropy and energy. 
We treat these composition rules as independent, including this way the purely entropic and
purely energetic non-additivity as particular cases.
At the end of this process we aim at having clear constraints
on the interplay between different non-extensivity parameters in a heterogeneous
thermal equilibrium.

By carrying out this program, we regard the problem as a
{\em purely thermodynamic question}, and elaborate the answer without relying on 
statistical arguments. Instead of additivity we consider general relations for fundamental 
thermodynamic quantities of composed systems. We refer to these relations as 
\textit{composition rules}. We accept the Maximum Entropy Principle and investigate the 
most general form of composition rules that are compatible with the factorizability of the 
equilibrium condition. 



The basic variables of classical thermodynamics are the so called extensive physical  
quantities (\(X_{}\)), like the energy ($E$), particle number of chemical components ($N_i$)  
and entropy ($S$). 
The adjective ''extensive'' denotes two different characteristics of these fundamental physical quantities. 
It is customary to assume that these quantities are added if we put two thermodynamic bodies 
together:
\begin{equation}
        X_{12}=X_{1}+X_{2}.
\end{equation}

This special property of composition is called \textit{additivity}. On the other hand traditionally one 
assumes that these quantities characterize the systems down to the smallest meaningful scale operating with finite densities of the extensive quantities \cite{Mac92b}:
\be
 \rho_X = \lim_{N\rightarrow\infty} \frac{1}{N}\sum_{i=1}^N X_i < \infty.
\ee{ext}  
Here we divided the body to \textit{N} different parts and \(X_i\) belongs to the  i-th of them.
This property is called \textit{extensivity}.
The two properties are related, but not equivalent. If a quantity is additive, then it is extensive, 
but there are  extensive and yet non-additive quantities, too  \cite{Bir08a,Tsa09b}.

Another important fundamental group of thermodynamic quantities are called intensives, 
and their definition\ is related to thermodynamic equilibrium and to 
the Zeroth and the Second Laws of thermodynamics. As a first step we do not analyze the 
relation of these concepts in their outermost generality, 
but rather start with a simple approach. We rely on the principle that  
the entropy is a function of the fundamental quantities and it is maximal in equilibrium. 
In this way we implicitly assume that the entropy is meaningful also (slightly) out of 
equilibrium. 
Let us consider the skeleton example, 
where there are only two thermodynamic bodies, both characterized 
solely by their respective energies \(E_1\) and \(E_2\).
By assuming that the energy is additive, we prescribe that the  total energy of the composed system 
of the two bodies \(E_{12}\) is given by the formula
\begin{equation}
E_{12}(E_1, E_2)=E_1+E_2.
\end{equation}

The bodies are characterized by their respective entropies and we assume that the entropy is additive:
\begin{equation}
        S_{12}(E_1, E_2)=S_1(E_1)+S_2(E_2)
\label{addent}
\end{equation}
with $S_i(E_i)$ being the respective equations of state. 
Maximizing the entropy, while assuming that 
the total energy of the two bodies is conserved (\(dE_{12}=0\)), although an exchange of energy between the
bodies is possible, one easily derives, that 
\begin{equation}
        dS_{12}(E_1,E_2) = \pd{S_1}{E_1}dE_1+\pd{S_2}{E_2}dE_2 = \left(S'_{1} -S'_{2}\right) \, dE_1 = 0,
\end{equation}
where the prime denotes the derivative of a single variable function. 
From this, one concludes that the equilibrium requires the equality of the derivatives:
\be
S'_{1}(E_1) =S_{2}'(E_2).
\ee{baseeq}

The Zeroth Law law of thermodynamics is  formulated as the requirement of the \textit{transitivity} of 
the thermodynamic equilibrium state. 
This transitivity property implies the existence of the 
empirical temperature. One can  easily see that  transitivity is satisfied by the above request 
and the  derivatives of the single body entropies are suitable empirical temperatures, because 
they are characteristics of the respective bodies and {\em do not depend on the properties of the 
partner body or the interaction}. Furthermore any monotonic function of the single body entropy 
derivative is an equally usable empirical temperature.

In this classical train of thoughts the Maximum Entropy condition 
written with additive composition rules leads to  equation \re{baseeq} 
that is \textit{factorizable}:\ the two sides depend on the 
quantities of the respective bodies. In the above case this property is trivial, 
and one realizes  that the additivity property of the energy and the entropy are 
important ingredients. In this way additivity
is a sufficient condition for the Zeroth Law, but the very question is whether it is also necessary.
 


We investigate now the problem of factorizability of the Zeroth Law when
both the entropy and energy follow non-additive composition rules. The Maximum Entropy Principle
with fixed combined energy leads to vanishing total differentials:
\ba
dS_{12} &= \pd{S_{12}}{S_1} S^{\prime}_1 dE_1 + \pd{S_{12}}{S_2} S^{\prime}_2 dE_2  =& 0
\nl
dE_{12} &= \pd{E_{12}}{E_1}  dE_1 + \pd{E_{12}}{E_2}  dE_2  =& 0.
\ea{TOT_DIFFS_ARE_ZERO}
Here we assumed that the composition laws for entropy and energy (and other further extensives)
are independent, i.e. the following natural assumption about the functional form of the composition rules
has been made: $S_{12}=S_{12}(S_1,S_2)$ and $E_{12}=E_{12}(E_1,E_2)$ are functions of the corresponding
variables of the subsystems for all possible equations of state
$S_1(E_1)$ and $S_2(E_2)$. 

A nontrivial solution for the energy changes $dE_1$ and $dE_2$ exists if the determinant
vanishes. This is given if
\be
\pd{E_{12}}{E_2} \cdot \pd{S_{12}}{S_1} S^{\prime}_1  
        =  \pd{E_{12}}{E_1} \cdot \pd{S_{12}}{S_2} S^{\prime}_2.
\ee{DET}
The most general form of partial derivatives now may include different two-variable functions
(indexed with $1$ and $2$):
\ba
 \pd{S_{12}}{S_1} &=& F_1(S_1) G_2(S_2) H_1(S_1,S_2),
\nl
 \pd{S_{12}}{S_2} &=& F_2(S_2) G_1(S_1) H_2(S_1,S_2),
\nl
 \pd{E_{12}}{E_1} &=& A_1(E_1) B_2(E_2) C_1(E_1,E_2),
\nl
 \pd{E_{12}}{E_2} &=& A_2(E_2) B_1(E_1) C_2(E_1,E_2).
\ea{PARC_DERIV_FACTORS}
The vanishing determinant condition (\ref{DET}) requires
\be
 A_2B_1C_2 \cdot F_1G_2H_1 S_1^{\prime} = A_1B_2C_1 \cdot F_2G_1H_2 S_2^{\prime}.
\ee{FACTORS_FACTORS}
This equation factorizes to $(E_1,S_1)$ and $(E_2,S_2)$ dependent terms only if
\be
\frac{C_2(E_1,E_2)}{C_1(E_1,E_2)} =  \frac{H_2(S_1,S_2)}{H_1(S_1,S_2)}.
\ee{GENERAL_FACTOR_CONDITION}
If one considers that the factorizability of the Maximum Entropy Principle should not
depend on any particular form of the equation of state, $S(E)$, then the above ratio
can only be a constant \footnote{This is analogous to the separable wave function ansatz in the
quantum mechanical description of the H-atom.}. 
Its value can easily be absorbed into one of the factorizing component
functions, so as an immediate consequence
\ba
 C_1(E_1,E_2) = C_2(E_1,E_2),
\nl
 H_1(S_1,S_2) = H_2(S_1,S_2).
\ea{MUST_HOLD}
These equalities are the basis for considering formal logarithms for the entropy and
energy separately. The factorized form of the Zeroth Law in this case,
\be
\frac{B_1 F_1}{A_1 G_1} S^{\prime}_1 = \frac{B_2 F_2}{A_2 G_2} S^{\prime}_2,
\ee{ZEROTH_LAW_FACTOR_FORM}
selects out the following choice for the thermodynamic temperature
\be
 \frac{1}{T} = \frac{B(E)F(S)}{A(E)G(S)} S^{\prime}(E).
\ee{TEM_GENERAL}
Finally, using the definitions
\ba
 \hat{L}(S) := \int \frac{F(S)}{G(S)} dS,
\nl
 L(E) := \int \frac{A(E)}{B(E)} dE,
\ea{FORM_LOG_DEFS_BOTH}
we arrive at
\be
 \frac{1}{T} = \pd{\hat{L}(S)}{L(E)}.
\ee{GENERAL_TEM}
The Zeroth Law requires that this common value is to be introduced as the reciprocal temperature.

The functions of the original thermodynamical variables, $\hat{L}(S)$ and $L(E)$
defined in eq.(\ref{FORM_LOG_DEFS_BOTH}) can be used to map the original composition
rules to a simple addition. Namely due to eqs.(\ref{PARC_DERIV_FACTORS}) and (\ref{MUST_HOLD}) 
one has
\ba
C_1 = \frac{1}{A_1B_2} \pd{E_{12}}{E_1} &=& \frac{1}{A_2B_1} \pd{E_{12}}{E_2} = C_2,
\nl
H_1 = \frac{1}{F_1G_2} \pd{S_{12}}{S_1} &=& \frac{1}{F_2G_1} \pd{S_{12}}{S_2} = H_2,
\ea{MUST_HOLD_FORM_LOG}
which can be re-arranged into the following form 
\ba
\frac{B_1}{A_1} \pd{E_{12}}{E_1} &=& \frac{B_2}{A_2} \pd{E_{12}}{E_2},
\nl
\frac{G_1}{F_1} \pd{S_{12}}{S_1} &=& \frac{G_2}{F_2} \pd{S_{12}}{S_2}.
\ea{ENTROPY_COMPOSITION}
Utilizing now the definitions (\ref{FORM_LOG_DEFS_BOTH}) both for $E_1,E_2$
and $S_1,S_2$ separately, the partial derivatives simplify:
\ba
  \pd{E_{12}}{L_1} &=& \pd{E_{12}}{L_2},
\nl
\pd{S_{12}}{\hat{L}_1} &=& \pd{S_{12}}{\hat{L}_2}.
\ea{FORM_DERIV}
The general solution of such partial differential equations is an arbitrary function of
the sum of variables:
\ba
 E_{12} = \Phi(L_1+L_2),
\nl
S_{12} = \Psi(\hat{L}_1+\hat{L}_2).
\ea{SUM_OF_VARIABLES}
In the still quite general case when the $\Phi$ and $\Psi$ functions are strict monotonic,
they are invertible. This inverse can be indexed by the composite system, so we arrive at
\ba
L_{12}(E_{12}) = L_1(E_1) + L_2(E_2),
\nl
\hat{L}_{12}(S_{12}) = \hat{L}_1(S_1) + \hat{L}_2(S_2).
\ea{CONCLUSION}
Since the $L_i(X_i)$ functions map non-additive quantities to the addition,
they can be called {\em formal logarithm} with right.
Since $\hat{L}(S)$ and $L(E)$ are additive quantities, they are also extensive.
Therefore eq.(\ref{GENERAL_TEM}) defines $1/T$  as a truly intensive quantity since it is a derivative of an extensive quantity with respect to an other  extensive quantity.

The extensively studied non-extensive entropy with additive energy composition rule
and its reverse, i.e. considering a non-additive energy while additive entropy, are
particular cases of the above result. Once a composition rule is the addition itself,
the corresponding formal logarithms become the respective identity functions, 
$L(E)=E$ and $\hat{L}(S)=S$. The classical result is recovered when both quantities
are composed additively.
For the entropic compositon rule Abe has derived the most general functional form
based on the factorization of the Zeroth Law in homogeneous equilibrium \cite{Abe2001}.
His result (cf. eq.$34$ in \cite{Abe2001}) conforms ours (\ref{CONCLUSION})
after the following mapping of the pseudoadditivity rule to the addition: 
\be
 \hat{L}(S)= \frac{1}{\lambda} \ln \left(1 + \lambda H_{\lambda}(S) \right).  
\ee{ABE_AND_US}


Physically, regarding different pieces of the \textit{same} material, all $L(E)$ and $\hat{L}(S)$ 
functions are the same -- this is the case of composition rules with a formal logarithm,
already proved to emerge in the limit of infinite repetitions of an arbitrary rule \cite{Bir08a}.
Statistics in the generalized thermodynamical limit of repeated compositions
therefore {\em ensures} the fulfillment of the Zeroth Law.
In a general setting, however, these functions - or ''only'' their parameters --
may differ. Moreover the mapping function for the interacting composed system
also may differ from both subsystems' corresponding functions.

On the one hand this leads to a variety of equilibria depending on the types of
the subsystems and composite systems (extensive or non-extensive). On the other hand
simple composition laws for homogeneous cases -- when all formal logarithm functions
are the same with the same value of all parameters -- may become more involved
when considering a Zeroth Law compatible heterogeneous equilibrium between different systems.
Our above results, eq.(\ref{CONCLUSION}) and eq.(\ref{GENERAL_TEM}), answer also the 
long debated question of equilibrium between extensive and non-extensive systems.

\section{The Zeroth Law: Transitivity}

The classical formulation of the Zeroth Law in thermodynamics emphasizes
the transitivity of the equilibrium condition without introducing the concept of entropy  \cite{HumThe11l}: when the subsystems $1$ and $2$ are in thermal equilibrium and independently
the subsystems $2$ and $3$, then {\em it follows} that also the subsystems
$1$ and $3$ are. This is a universal principle.

Our above condition, derived from factorization of the constrained Maximum Entropy 
Principle, automatically satisfies this transitivity. We demonstrate this on the
use of non-additive energy composition rule, other cases can analogously be derived.
The key observation is that we have established additivity of composite functions of
the energies of the respective subsystems, $L_i(E_i)$. Therefore it is natural
to assume that these functions are characteristic to the subsystems and only the
double-indexed formal logarithms, $L_{ij}(E_{ij})$, are characteristic to the
interaction between subsystems in equilibrium. In this way all subsystems develop
the same individual formal logarithm irrespective to which other system
they equilibrate with.


Assuming namely the opposite, i.e. a partner-dependent individual formal logarithm,
the transitivity would be violated. Let us regard the three possible pairings of three subsystems. The composite energies satisfy
\ba
 E_{12} &=& \Phi_{12}(L_1(E_1)+L_2(E_2)),
\nl
 E_{23} &=& \Phi_{23}(\tilde{L}_2(E_2)+L_3(E_3)),
\nl
 E_{13} &=& \Phi_{13}(\tilde{L}_1(E_1)+\tilde{L}_3(E_3)).
\ea{THREE_COMPOSIT}
If $\tilde{L} \ne L$, then the equilibrium condition is not automatically transitive. 
In the special case of additive entropy but non-additive energy composition 
considered here equation (\ref{DET}) reads pairwise as
\ba
 S_1^{\prime}(E_1) \pd{E_{12}}{E_2} &=& S_2^{\prime}(E_2) \pd{E_{12}}{E_1} 
\nl
 S_3^{\prime}(E_3) \pd{E_{23}}{E_2} &=& S_2^{\prime}(E_2) \pd{E_{23}}{E_3} 
\nl
 S_1^{\prime}(E_1) \pd{E_{13}}{E_3} &=& S_3^{\prime}(E_3) \pd{E_{13}}{E_1}. 
\ea{TRANSITIV_ZERO_COND}
From here the ratios of respective $S^{\prime}(E)$ factors can be expressed, 
\ba
  \frac{S^{\prime}_1(E_1)}{S^{\prime}_2(E_2)} &=& \frac{\pd{E_{12}}{E_1}}{\pd{E_{12}}{E_2}} 
\nl 
  \frac{S^{\prime}_2(E_2)}{S^{\prime}_3(E_3)} &=& \frac{\pd{E_{23}}{E_2}}{\pd{E_{23}}{E_3}} 
\nl 
  \frac{S^{\prime}_1(E_1)}{S^{\prime}_3(E_3)} &=& \frac{\pd{E_{13}}{E_1}}{\pd{E_{13}}{E_3}},
\ea{THREE_RATIOS}
leading to the condition
\be
 \frac{ \pd{ E_{12} }{ E_1 }  }{ \pd{ E_{12} }{ E_2 }  } \, 
 \frac{ \pd{ E_{23} }{ E_2 }  }{ \pd{ E_{23} }{ E_3 }  }  = 
 \frac{ \pd{ E_{13} }{ E_1 }  }{ \pd{ E_{13} }{ E_3 }  }.
\ee{TRANSTIV_PDE}
This compared with the form (\ref{THREE_COMPOSIT}) reveals the following consistency
requirement:
\be
\frac{ \Phi^{\prime}_{12} L^{\prime}_1 }{ \Phi^{\prime}_{12} L^{\prime}_2 } \,
\frac{\Phi^{\prime}_{23} \tilde{L}^{\prime}_2}{\Phi^{\prime}_{23} {L}^{\prime}_3}  =
\frac{\Phi^{\prime}_{13} \tilde{L}^{\prime}_1}{\Phi^{\prime}_{13} \tilde{L}^{\prime}_3}. 
\ee{THREE_CONSISTENT}
This condition can be reduced easily to obtain
\be
\frac{L^{\prime}_1}{\tilde{L}^{\prime}_1} =
\frac{L^{\prime}_2}{\tilde{L}^{\prime}_2} \,
\frac{L^{\prime}_3}{\tilde{L}^{\prime}_3}. 
\ee{THREE_REDUCED}
Since this equality is required for any permuted arrangement of the indices $1$, $2$ and $3$,
one concludes that the transitivity of thermal equilibrium can only be satisfied
if the $\tilde{L}^{\prime}$ and $L^{\prime}$ functions are identical. 
This is a necessary and sufficient condition.
Adding the observation that a physically sensible composition rule satisfies the
triviality condition, i.e. a composition with zero does not change the value,
in general $L(0)=0$ is required. In this case also the $\tilde{L}$ and $L$ functions
themselves are identical. Requiring furthermore that for small energies --
when non-additive effects are, as a rule, relatively reduced -- the addition re-emerges,
also $L^{\prime}(0)=1$ is set. In the following discussion we assume that these
properties are fulfilled.

\section{The Example of Tsallis' Entropy Composition Formula}

A non-additive (and for factorizing probabilities also non-extensive) entropy formula,
promoted by Tsallis \cite{Tsa09b,Tsa88a}, satisfies the following composition rule
\be
S_{12} = S_1 + S_2 + \hat{a} S_1 S_2.
\ee{TSALLIS_COMBO}
Here we used the shortening notation $\hat{a}=1-q$. The additive entropy systems
have $\hat{a}=0$, non-extensive systems another value of this parameter.
The question, how two such systems with different $\hat{a}$ parameters can come into
thermal equilibrium compatible with the Zeroth Law, was raised on quite a few
occasions \cite{Nau03a,Wan02a}. 

According to our main result in this paper the answer is affirmative for
a Zeroth Law compatible equilibrium, provided that the general case is treated
via the formal logarithm. The formal logarithm for the above rule is easy to derive
by observing that
\be
1 + \hat{a} S_{12} = 1 + \hat{a}S_1 + \hat{a}S_2 + \hat{a}^2 S_1 S_2 
        = (1+\hat{a}S_1) \, (1+\hat{a}S_2).
\ee{TSALLIS_EASY}
The product is mapped to the addition by the logarithm and scaled to satisfy
$\hat{L}^{\prime}(0)=1$:
\be
\hat{L}(S) = \frac{1}{\hat{a}} \ln (1+\hat{a}S).
\ee{TSALLIS_FORMAL_LOG}
This trivial property has been observed several times, see e.g Ref.\cite{Abe01a}.
For a general thermal equilibrium between two subsystems with different values
of the non-additivity parameter, $\hat{a}_1$ and $\hat{a}_2$ a Zeroth Law compatible
equilibrium emerges - according to eq. \re{ENTROPY_COMPOSITION} - if
\be
\frac{1}{\hat{a}_{12}} \ln (1+\hat{a}_{12}S_{12}) =  
        \frac{1}{\hat{a}_1} \ln (1+\hat{a}_1S_1) +  \frac{1}{\hat{a}_2} \ln (1+\hat{a}_2S_2).
\ee{OUR_RESULT}
Here one observes that a further non-extensivity parameter, $\hat{a}_{12}$, was also introduced.
It is necessary for counting for non-additivity in the composed systems between
elements (atoms, particles) of the respectively different subsystems. 
See \cite{BirPur08a} for a parton cascade based numerical simulation 
with a non-additive energy composition rule. 
The above relation (\ref{OUR_RESULT}) leads to the direct expression
\be
S_{12} = \frac{1}{\hat{a}_{12}} \left( (1+\hat{a}_1S_1)^{\hat{a}_{12}/\hat{a}_1} \, 
        (1+\hat{a}_2S_2)^{\hat{a}_{12}/\hat{a}_2}  -1 \right).
\ee{OUR_S12_DIRECT}
In case of homogeneously non-additive sub- and composed systems one considers 
$\hat{a}_{12}=\hat{a}_1=\hat{a}_2=\hat{a}$ and obtains equation (\ref{TSALLIS_COMBO}). 
We note here that the composition formula (\ref{OUR_S12_DIRECT}) does not fulfill
the naively expected property when composing a finite entropy and a zero entropy subsystem, 
i.e. $S_{12}(S_1,0)\ne S_1$ and $S_{12}(0,S_2)\ne S_2$. This property is, however,
trivially satisfied for $\hat{L}_{12}(S_{12})$ when one of the subsystem formal logarithms,
$\hat{L}_1(S_1)$ or $\hat{L}_2(S_2)$, vanishes.

For a composition of two non-additive subsystems in general one obtains
\be
S_{12} = \Psi_{12} \left( \hat{L}(\hat{a}_1,S_1) + \hat{L}(\hat{a}_2,S_2) \right).
\ee{AHA}
In case of $\hat{a}_1$=0, i.e. considering thermal equilibrium between an additive
and a non-additive system, one achieves 
\be
S_{12} = \Psi_{12} \left( S_1 + \hat{L}(\hat{a}_2,S_2) \right).
\ee{MIXED_AHA}
From the viewpoint of Zeroth Law compatibility it is strongly advised to use the
additive formal logarithm of any non-extensive entropy formula. In case of the
Tsallis entropy,
\be
S_T = \frac{1}{\hat{a}} \left( \sum_i\limits p_i^{1-\hat{a}} - p_i \right),
\ee{TSALLIS_FORMULA}
with $\hat{a}=1-q$ and the normalization $\sum_i p_i = 1$, its formal logarithm turns out to be
the well known R\'enyi entropy \cite{Ren61a}:
\be
S_R = \hat{L}(\hat{a},S_T) = \frac{1}{\hat{a}} \ln (1+\hat{a}S_T) = \frac{1}{1-q} \ln \sum_i\limits p_i^{q}.
\ee{RENYI_FORMULA}

\section{Equilibrium distribution function}

Based on the above arguments the general thermal equilibrium state satisfying
$\beta = 1/T = \partial \hat{L}(S)/\partial L(E)$ motivates us to maximize
$\hat{L}(S)-\beta L(E)$ when looking for canonical energy distributions \cite{Bir11b}.
Since the formal logarithms are additive -- even if the original quantities to be composed
were not -- a distribution can be obtained from
\be
  \hat{L}(S)\left[ p_i \right] - \beta \sum_i p_i L(E_i) -\alpha \sum_i p_i = {\rm max}.
\ee{GENERAL_CANONICAL}
For example for leading order non-additivity according to the composition rule
of type (\ref{OUR_S12_DIRECT}) the R\'enyi entropy, 
\(\hat L(S)=\frac{1}{\hat{a}} \ln  \sum_i p_i^{1-\hat{a}} \), is to be maximized. 
Denoting the formal logarithm parameter for the entropy composition by $\hat{a}=1-q$ 
and for an analogous energy composition by $a$, one considers
\be
\frac{1}{\hat{a}} \ln  \sum_i p_i^{1-\hat{a}}  - \beta \sum_i p_i \frac{1}{a} \ln(1+aE_i)
- \alpha \sum_i p_i = {\rm max}.
\ee{RENYI_CANONICAL}
The extremal condition results in 
\be
  p_i =A\left(b(\alpha + \beta L_i)\right)^{-\frac{1}{\hat a}}
\ee{edist}
Here we have introduced the notation
\be
A = e^{-\hat L(S)}, \qquad L_i =\frac{1}{a} \ln(1+aE_i), \qquad b=\frac{\hat a}{1-\hat a}.
\ee{}
Then the normalization, the average and  the definition of the entropy read as
\ba
1 &=& A \sum_i  \left(b(\alpha + \beta L_i)\right)^{-\frac{1}{\hat a}},\\
\left<L\right> &=& A \sum_i L_i \left(b(\alpha + \beta L_i)\right)^{-\frac{1}{\hat a}},\\
A^{-\hat a} &=& A^{1-\hat a} \sum_i  \left(b(\alpha + \beta L_i)\right)^{1-\frac{1}{\hat a}}. 
\ea{}
From this set of equations one obtains the condition
\be
1=b\alpha+b\beta\left<L\right>. 
\ee{maincond} 
Therefore the equilibrium distribution simplifies to 
\be
 p_i = A \left(1+b\beta(L_{i}-\left<L\right>) \right)^{-1/\hat{a}} = 
 \frac{1}{Z} \left(1+\hat{a}\hat\beta L_i \right)^{-1/\hat{a}} =
 \frac{1}{Z} \left(1+\hat{a}\hat{\beta}\frac{1}{a} \ln(1+aE_i) \right)^{-1/\hat{a}}.
\ee{RENYI_DISTRIB}
Here we have introduced the shorthand notations: 
\be
Z = \frac{1}{A } \, (1-b \beta \left<L\right> )^\frac{1}{\hat a}, \qquad 
\hat{\beta} = \frac{\beta}{1-\hat{a}(1+\beta \left<L\right>)}.
\ee{cons1}
We should keep in mind that the reciprocal temperature, distinguished by the Zeroth Law, 
is the Lagrange multiplier \(\beta\). This is reflected well by the whole formalism, 
because the usual thermodynamic relations are valid. A distribution similar to 
(\ref{RENYI_DISTRIB}) - but with negative values of the energy non-additivity parameter $a$ - 
has been  derived for particle energy spectra inside jets, 
observed in  electron-electron collision experiments, 
by taking into account multiplicity fluctuations \cite{UrmEta11m}. 

It is enlightening to consider now cases, where one or the other quantity is
composed by additive rules.
In the limit of additive entropy but non-additive energy composition rules 
(\(\hat a \rightarrow 0 \)) the canonical distribution approaches
\be
 p_i = \frac{1}{Z_{0}} \left( 1+aE_i\right)^{-\beta/a},\qquad
 \text{where} \qquad \ln  Z_0 = S_{BG} - \beta \left<E\right>, 
\ee{ENTR_ADD_ENERG_NON}
and \(S_{BG}\) is the Boltzmann-Gibbs entropy. 
For non-additive entropy and additive energy on the other hand a differently parametrized
power-law tailed distribution emerges:
\be
p_i = \frac{1}{Z} \left( 1+\hat{\beta} \hat{a} E_i\right)^{-1/\hat{a}}.
\ee{ENERG_ADD_ENT_NON}
How to distinguish between these two approaches in a physical situation?
In some cases the entropic non-additivity, in some other cases the energetic non-additivity
may be preferred. To mention an example, the ideal gas of massless particles can be
considered. The average energy of a single particle, $\exv{E}$, 
known to be connected to the temperature,
leads to different formulas in these two cases. For the entropic non-additivity one obtains
the familiar result,
\be
T =  \frac{1}{f} \exv{E}, 
\ee{TEM_ENTROP}
with $f=D$ for massless particles in $D$ spatial dimensions and with $f=D/2$ for
the non-relativistic energy relation $E=\vec{p}^2/2m$. This result is
independent of the value of the entropic non-additivity parameter $\hat{a}$.
For the general case using $E= c \sqrt{\vec{p}^2+(m c)^2}$, however, the result will
depend on the non-additivity parameter \cite{Lavagno,Alberico}.

On the other hand for the energetic non-additivity, applied under the above conditions, one gets
\be
T = \frac{\exv{E}}{f+(f+1)a\exv{E}}.
\ee{TEM_ENER}
In this case for $a > 0$ a maximal (so called {\em limiting}) temperature arises: $T \le 1/(f+1)a$. 
We note by passing that fixing 
$\exv{ L(a,E)}$ instead, the limiting temperature is $T \le 1/fa$ \cite{BirPes06a}. We would like to emphasize again, that this \(T\) is the thermodynamical temperature - being intensive and satisfying the Zeroth Law - as defined in eq. \re{GENERAL_TEM}.

\begin{figure}
\begin{center}
        \includegraphics[width=0.66\textwidth,angle=-90]{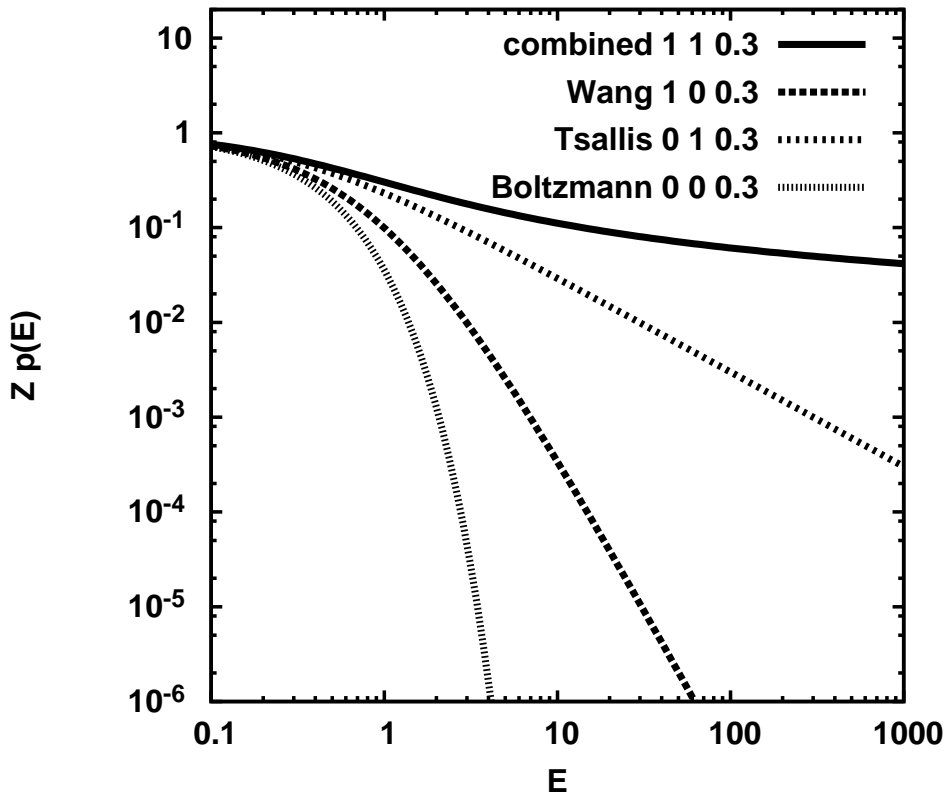}
\end{center}
\caption{ \label{Fig1}
   The stationary energy distributions for different pairings of additive
   and non-additive systems with relativistic dispersion relation and
   by using the formal logarithms $L(E)=\ln(1+aE) / a$ and  $\hat{L}(S)=\ln(1+\hat{a}S)/\hat{a} $.
   In the legend the parameters are $a$, $\hat{a}$ and $\hat{T}=1/\hat{\beta}$, in this order.
}

\end{figure}

In  Figure \ref{Fig1} stationary energy distributions are depicted
for the equilibrium between additive systems, between energetically or entropically
non-additive ones, and for the case when both non-additivities are considered.
They are labeled according to the non-additivities as "Boltzmann", 
"Tsallis", "Wang" and "combined" respectively.
The parameters given in the legend are $a$, $\hat{a}$ and 
$\hat{T}=1/\hat{\beta}$ in this order, in the corresponding energy units.


\section{Summary and conclusions}

In this paper we have reversed the usual approach to  Zeroth Law of thermodynamics in 
non-extensive thermostatistics. Instead of assuming some particular non-additive entropy 
form or composition rule for either the entropy or other basic, traditionally extensive, quantities,
we have regarded the Zeroth Law more fundamental, and derived consequent requirements for the 
general composition rules. 
In our analysis we have considered non-additive composition rules
both for the energy and entropy while composing two subsystems. In principle this
method can be extended to all extensive thermodynamical variables.

We have proved that the Zeroth Law together with the Maximum Entropy Principle strictly restricts  
the possible functional form of the composition rule, and this restriction can be resolved
in a simple manner by using formal logarithms both for the energy and entropy functions.
{\em The Zeroth Law essentially enforces the additivity of these formal logarithms
of the basic extensive quantities, but not the additivity of the quantities themselves.}
In this way it is true that the Zeroth Law enforces additivity, but it is also true that
not only the Boltzmannian formula is compatible with the basic principles of thermodynamics.

The temperature, defining thermal equilibrium among systems following such
generalized composition rules instead of the simple addition, 
is the inverse of the  Lagrange multiplier associated to formal logarithms of the energy 
and entropy in the  Maximum Entropy Principle. 

Finally we briefly outlined some important properties of the arising 
canonical equilibrium distributions. In the general case, compatible to the Zeroth Law,
these are not at all restricted to a Gibbs-exponential of the individual energy.

In our approach several, at the first sight seemingly paradoxical, aspects of non-extensive 
thermostatistics can be explained. We list the most important ones of them.
\begin{enumerate}
\item The general form of the entropy composition rule \re{OUR_S12_DIRECT}, 
        derived from the Tsallis like composition \re{TSALLIS_COMBO}, reveals several $q$ parameters. 
   The use of the corresponding formal logarithm with all the same $q$ parameter
   and functional form describes the special case of {\em homogeneous} equilibrium.
   According to our analysis of the factorizable form of the Zeroth Law,
   however, formal logarithms with different functional forms and parameters
   also can be used to satisfy all requirements. This describes a {\em heterogeneous}
   equilibrium; among others between extensive and non-extensive systems.
A practical example is given by a Monte
Carlo simulation of non-extensive systems consisting of 'red' and 'blue' particles. The energetic non-additivity parameters \(a_{1}\) for the red-red interaction, \(a_2\) for the blue-blue and \(a_{12}\) for the red-blue one all can differ.
\item  Heterogeneous composition formulas can be derived to any proposition for an entropy formula by determining
   the corresponding formal logarithm.

\item It is possible to introduce Zeroth Law compatible energy or entropy non-additivity 
   either separately or simultaneously. Contrary to previous expectations \cite{Wan02a,Sca10a}, 
   the two different non-additivities do not imply each other.


\item Either certain energy or entropy non-additivities may lead to power law-tailed
  equilibrium distributions, but other properties, like the equipartition law, clearly distinguish 
  these two cases.
  For example a limiting temperature is the consequence of a repulsive energetic non-additivity of type
  $E_{12}=E_1+E_2+aE_1E_2$, but an entropic non-additivity of similar form does not result in such a property.

\item  As it was emphasized in the introduction and in \cite{Tsa09b}, additivity is a sufficient but 
  not a necessary condition for extensivity ( cf. eq. \ref{ext}). 
  Some non-additive rules, when repeated many times
  on re-scaled systems, may lead to the addition asymptotically \cite{Bir08a}.
  It is interesting to note that requiring Zeroth Law compatibility for general composition rules
  leads to the conclusion that these rules should also be \textit{associative}. 
  E.g. for the energy composition associativity requires
  \be
        E_{12,3}(E_{12}(E_1, E_2),E_3)=E_{1,23}(E_1,E_{23}(E_2, E_3)).
  \ee{ASSOC}
  Utilizing the representation of the composition rule with correspondingly indexed 
  formal logarithms, one easily derives that
  \be
        L_{12,3}(E_{12,3}) = L_1(E_1)+L_2(E_2)+L_3(E_3)  = L_{1,23}(E_{1,23})
  \ee{ASSOC_LOGS}
  i.e. the associativity of the formal logarithms of the different level composed subsystem energies
  has to be satisfied. 

\item Our approach to extensivity and additivity is also different from a recent classification and use 
  of different non-additive entropies and related concepts  (\cite{Tsa09b} p.91-106). 
  All our results are  not bound to a particular composition rule, like \re{TSALLIS_COMBO}, 
  or to purely entropic non-additivity. Any of the traditionally extensive quantities may be composed
  by almost arbitrary non-additive rules; all these versions of thermodynamics are Zeroth Law
  compatible if the Maximum Entropy Principle is formulated by using all over the corresponding
  formal logarithms.
  In particular our conclusion is that in thermal equilibrium the R\'enyi entropy, 
  as the formal logarithm to the Tsallis entropy, 
  is to be maximized with the traditional normalization of probabilities.   
 \end{enumerate}

Finally we note that our analysis and statements are restricted to thermal equilibrium.
Far from the equilibrium state there may be other properties, like the convexity, which
could indicate a preference for using non-additive formulas (cf. the comparison between
Tsallis- and R\'enyi-entropy in Ref. \cite{Tsa09b}).

\section{Acknowledgment}

The work was supported by the Hungarian National Science Research Fund OTKA (K68108, K81161).

\bibliographystyle{unsrt}

\end{document}